# Multi-modal Fusion using Fine-tuned Self-attention and Transfer Learning for Veracity Analysis of Web Information

Priyanka Meel[1], Dinesh Kumar Vishwakarma[2]
Biometric Research Laboratory, Department of Information Technology
Delhi Technological University, New Delhi, India-110042

**Abstract**

The nuisance of misinformation and fake news has escalated many folds since the advent of online social networks. Human consciousness and decision-making capabilities are negatively influenced by manipulated, fabricated, biased or unverified news posts. Therefore, there is a high demand for designing veracity analysis systems to detect fake information contents in multiple data modalities. In an attempt to find a sophisticated solution to this critical issue, we proposed an architecture to consider both the textual and visual attributes of the data. After the data pre-processing is done, text and image features are extracted from the training data using separate deep learning models. Feature extraction from text is done using BERT and ALBERT language models that leverage the benefits of bidirectional training of transformers using a deep self-attention mechanism. The Inception-ResNet-v2 deep neural network model is employed for image data to perform the task. The proposed framework focused on two independent multi-modal fusion architectures of BERT and Inception-ResNet-v2 as well as ALBERT and Inception-ResNet-v2. Multi-modal fusion of textual and visual branches is extensively experimented and analysed using concatenation of feature vectors and weighted averaging of probabilities named as Early Fusion and Late Fusion respectively. Three publicly available broadly accepted datasets All Data, Weibo and MediaEval 2016 that incorporates English news articles, Chinese news articles, and Tweets correspondingly are used so that our designed framework's outcomes can be properly tested and compared with previous notable work in the domain. The accuracy varies on different datasets using different fusion methods. The recommended multi-modal technique accomplishes 97.19 % highest accuracy on All Data dataset using ALBERT & Inception-ResNet-v2 late fusion framework, which is notably better than all the previous works in the domain of veracity and credibility assessment of online information contents.

# 1. Introduction

In the present era, the accessibility and readiness of the internet have made it possible to get real-time news for people worldwide. Nevertheless, these ongoing developments in technology and the distribution of information through various networks, including social media have provocatively amplified the spread of rumors and fake news. The real problem is not disseminating phony details but the fact that people tend to accept whatever is fed to them through these online platforms. Hence, information pollution is one of the most prominent intimidations to ethical journalism, the autonomy of expression, and in a way, the fundamentals of democracy. Just because of the utter difficulty of identifying rumors, they disseminate like wildfire on online platforms, increasing the issue's severeness multiple folds. It is comparatively easy to identify fake service and product reviews by their tenor and intent; fake news can be highly elusive and deceptive. So obviously, it would be pretty naive to expect an average reader, who doesn't always have the time, energy or even the resources, to perform the tricky task of identifying fake news. The explosive growth and deadly repercussions of fake news have made it inevitable to find solutions that can wipe this problem out once and for all.

Though publishing and spreading false information is not something new, the usage of the term "fake news" is. It was first used only at the end of the 19th century in a newspaper, Cincinnati Commercial Tribune in the United States (US) [1]. Since then, many researchers and scholars [2], [3], [4] have tried to outline it in different ways such as: misinformation, disinformation, clickbait, hoax, propaganda, satire, fake news, rumor etc. But what is common in all their explanations is that fake news is a portion of the information presented to be accurate and real, but in reality, it has no facts to support it. In more cases than not, it is spread intentionally to achieve some hidden personal agenda. While many of these fabricated tales are not intended to be harmful and take the form of funny jokes and satire, others indeed have more disturbing implications. Because the accountability factor is missing, anyone in any part of the world can claim whatever he feels like without fearing the consequences, which practically don't exist. This presents a long list of problems, with the most critical of them being the damage to public mental health.

According to Dr. Chrysalis Wright, Associate Lecturer of Psychology at the University of Central Florida [5], the most lethal aspect of fake news is that it makes us only see one aspect of a story, which in fact may be completely fabricated. But it's so captivating that we don't always care about the complete picture. We get upset and emotional, making us want to share it with our friends and family. That's how the cycle begins. In no time, a small piece of fake news is getting circulated like wildfire worldwide.

Many people believe that the increase in fake news is because of the declining popularity of newspapers and other traditional media sources. The journalist's and social media blogger's carefree attitude of sharing stories without verifying them first adds severity to the issue multiple folds. The erosion of trust in traditional news sources created a big void or an opportunity for the fake news spreaders to unfortunately take advantage. Technological advancements have made it possible to spread misleading stories in every corner of the world. Fake news is being read, processed and forwarded before a person even gives a second to verify its authenticity. Fig. 1 highlights some of the widely circulated fake stories in recent times.

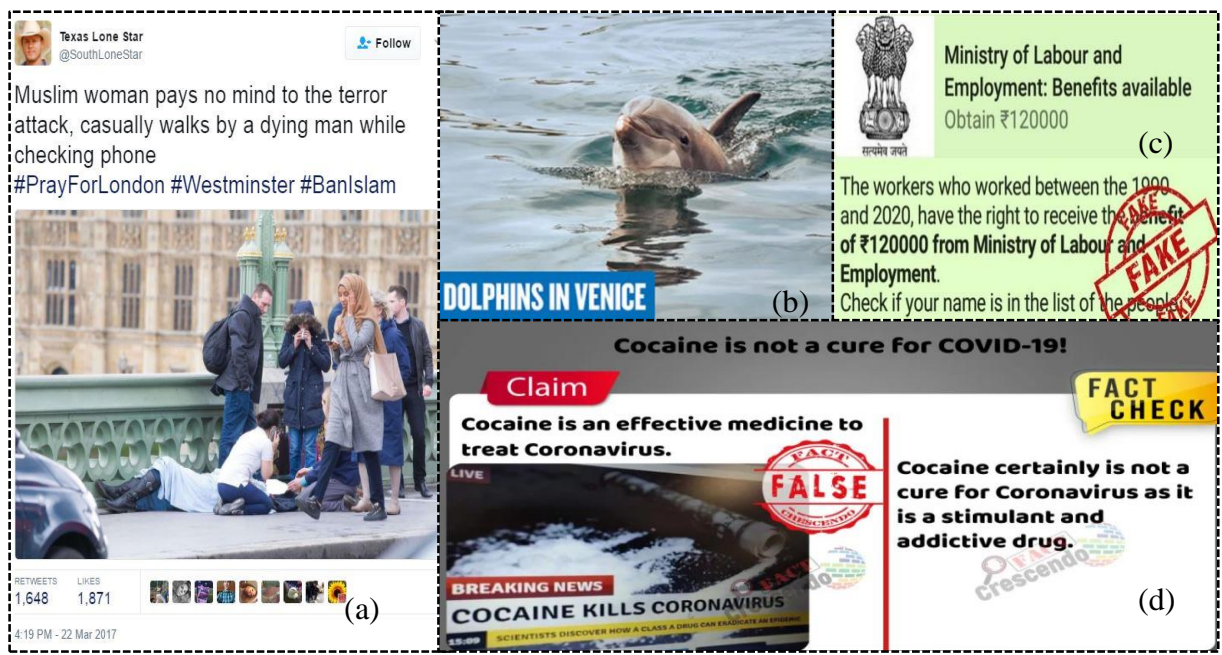

Figure 1: Instances of fake news on web platforms in multiple data modalities

On the surface, it might seem that fake news doesn't have any extreme consequences. But in reality, it has the potential to ruin the lives of people permanently. One such example is the 2017 terror outbreak in London [6] that happened on 22nd March 2017. Just hours after half a dozen people were killed and half a century got injured, an image of a middle-aged lady wearing a hijab and speaking on the phone on Westminster Bridge, the site of the attack, got viral all over the internet. Thousands of people shared it on their profiles claiming the woman, as a Muslim, was unsympathetic to the sufferings of fatalities around her, with the hashtag "#BanIslam" that got trending everywhere as in Fig 1(a). Later, the account which first tweeted the image was suspended after it was found to be a Russian bot and the image was taken off from the internet. But the mental damage it caused to the woman could never be undone. She issued a statement after being devastated by the negative attention she encountered due to this fake news and spoke about how this news has ruined her identity in the community in which she lives. That's how poisonous a small piece of fake news can be. It can spoil the lives of innocent people in a matter of hours.

During coronavirus lockdown, social media was flooded with animal news, healing nature claims and the images of dolphins in canals of Venice as in Fig 1(b). Sadly enough, the viral images were not of Venice canals, they are from a different region of Italy. But the question that arises is that why is fake news created in the first place. What could a person possibly gain from it? The most common and apparent reason is money. Fake news pays well. On various social media sites, the posts that trend and gather a higher number of views generate revenue. And what better way to get a viral post than to create something controversial and buzzy, something that can be catchy enough to garner a lot of viewers. And it's not only the person who is being targeted in the fake news but also the money that gets paid to the creator comes out of our pockets. Another widespread motive to generate fake news is political propaganda. Often during the elections, a sudden surge in the spread of fake news is observed. One case in point being the 2016 US elections, which got heavily swayed by the impact of fake news [7], [8]. Some political experts even claim that Donald Trump won't have won the elections if it weren't for the massive impact of fake news.

It is well-established by now that fake news is a grave issue that needs to be uprooted from its core. But the task of eradicating it is not as simple as it might look. Some of the significant challenges need to tackle to classify fake news from a real one can be listed as follows:

• The lack of a standard definition of fake news makes it tough to classify one. There's no exact way to know that where we should draw the line. It is a significant concern since it blurs the boundary between real and fake.

• Fake news spreads at lightning speed. By the time a piece of information is doubted for its authenticity, it has already reached thousands of people, feeding them with false claims and rumours.

• A fake news article's tone is not as explicit and direct as in a fake review or recommendation. Since the intent is so subtle and hidden, finding an alternate way of identifying fake news gets tricky.

• Sometimes, a piece of information is not entirely true or false. Some fake news creators mix

fake elements in true stories to misguide and influence people. This makes the whole task extremely complex since the information is not entirely fake but only to an extent.

Multimedia has become an integral part of human life as it has more conclusive, convincing and long-lasting memory effects. Almost every circulating news story is strengthened with accompanying images or videos. Multiple data formats have made forgery identification in online circulated news articles quite complicated because each data format has different characteristics. As the inherent attributes of varied data formats differ considerably, so as the techniques to detect their forgeries. Therefore, we propose the framework of multi-stream fusion to harness the independent benefits of the properties of textual and visual data streams, then merge them to tag the overall instance as real or fake. Transformer-based BERT [9] and ALBERT [10] language models, solely based on attention mechanism and bidirectionally trained to have a more profound sense of language context and flow, are used for textual feature extraction. Inception -Resnet-v2 [11] deep neural network is employed for training the classifier with the visual data. The effectiveness of multi-stream fusion is being explored by designing independent late and early fusion architectures. The work is expected to enrich the research community with the naïve framework of multi-modal fusion using self-attention and deep neural networks. The key contributions of the work are as follows:

- The problem of fraudulent content in text and image multi-modal data format is being identified and addressed for veracity analysis of web information contents.

- Textual feature extraction and model training is being done using BERT and ALBERT bidirectional self-attention-based transformers.

- Visual data is being analysed by harnessing transfer learning capabilities using fine-tuned Inception-ResNet-v2 deep neural network architecture.

- The proposed framework focused on two independent multi-modal fusion frameworks of BERT and Inception-ResNet-V2 as well as ALBERT and Inception-ResNet-V2.

- The multi-modal fusion of textual and visual branches is extensively experimented and analysed using concatenation of feature vectors and weighted averaging of probabilities named Early Fusion and Late Fusion respectively.

- Three publicly available broadly accepted datasets All Data, Weibo and MediaEval 2016 that incorporate English news articles, Chinese news articles, and Tweets correspondingly are used to test and compare our designed framework's outcomes with previous notable work in the domain.

The work is explained in its entirety under five sections and multiple subsections. Section 1 emphasis on the need of addressing the information pollution issue with few widely circulated fake news examples. Literature survey and research gap is being highlighted in section 2. Section 3 details in the technical description of the proposed architecture and its different components. Experimental settings, datasets, system specification, detailed analysis of obtained results and comparative study is being done in section 4. Eventually, section 5 summarizes the work and recommends the prospective directions for upcoming research. Our work's central objective is to propose an effective and efficient framework that can correctly classify fake news. It leverages advanced deep learning models to accomplish this task using both text and image features of the news data. We hope that this work will prove to be a landmark within the province of veracity analysis and encourage other researchers to propose further advancements that would take the scope of the solution to new heights.

## 2. Related Works

Information circulating online on web platforms has various attributes like headline/title, text/body, author, associated image, URL, etc. Any changes made in any of the features incline news to behave abnormally, making it a piece of fake news or a rumour. Researchers have tried to verify different attributes, webpage URL, author, text, associated image, time series analysis and propagation statistics to design models for veracity analysis. False contents, also popularly termed as information pollution or infodemic, are amplifying at an alarming rate. Efforts have already started for providing solutions to detect and mitigate fraudulent content using up-to-date artificial intelligence technologies. A considerable number of methodologies embrace fake news uncovering textual data. With the rise of multimedia data on users' posts and news, research incorporating the identification of visual fake news has increased and improved the preciseness of algorithms.

During the COVID-19 pandemic, the circulation of fake news or rumors is being highly observed. In an article titled "**Fake News, Real Arrests**" [12], the author categorically emphasized the increase in the number of arrests by police due to the spread of fake news across different states of India. It has also been quoted that "**Virus of fake news spreads faster than Corona Virus itself**". Social media is the most accessible tool to spread fake news rapidly and bots are being programmed to make the task easier. Existing literature enumerates several criteria for classifying the veracity analysis approaches. We organize varied methods available in literature according to the feature being analyzed to verify the truthfulness of the information. The following Fig 2 represents the categories of organization in our work.

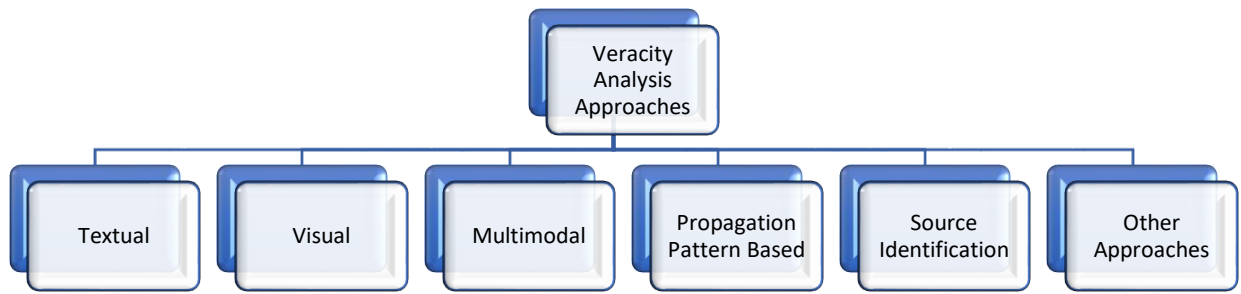

Figure 2: Classification of Veracity Analysis Approaches

## 2.1 Text-Based Approaches

Many researchers have exploited features from textual contents like textual frequency-based features, semantic features, sentiment analysis, writing style, text pattern, the polarity of contents etc. Pan et al. [13] have studied how the style of writing is responsible for shaping the views of the people. They used a unique structure called knowledge graphs to categorize given news text as fake or real. Gautam and Jerripothula [14] put forward a novel framework for text fake news detection using Random Forest classifier that leverages the capabilities of paraphrasing tool Spinbot, grammar checking tool Grammarly and Glove word-embedding for feature extraction. Li and Zhou [15] employ BERT language model for binary text fake news detection by connecting the dots between the claim and pieces of evidence. Lengthy text news articles are first summarized to extract the claims as well as keywords that are subsequently searched on web to extract the related articles to treat then as evidences or facts to be matched against the claim. Finally, the scores of similarities are fed into a fact verification model to classify a news article as real or fake. To detect ambiguous information on COVID 19 Elhadad et al. [16] collected factual data from WHO, UNICEF, United Nations and epidemiological data from different fact check websites to consolidate into a repository. This accumulated factual data is used to build a fake news classification system based on ensemble of ten different machine learning classifiers using seven feature extraction techniques implemented and tested against twelve performance metrices.

## 2.2 Visual Approaches

Images are an integral part of print, electronic and digital media news. Visual snapshots alone or mostly combined with text overwritten them when manipulated spread forgery to a great extent. A novel method is proposed [17] to detect photographic splicing detection via illumination inconsistencies and deep learning. This approach is capable enough of locating forged regions, eliminates the laborious feature engineering process and provides outperforming accuracies on the same datasets paralleled to state-of-the-arts. Elkasrawi et al. [18] devised a semi-automatic approach of verifying the credibility of images in online articles by means of a two phased process embraces clustering and hierarchical image feature analysis, accounts to highest 88% accuracy for a dataset of 50 images. Fake images generated by Generative Adversarial Networks (GANs) by changing context and semantics of original image via. Image-to-image translation looks very realistic. The study proposed by Marra et al. [19] performed on a dataset of 36302 images describes the performance of multiple image forgery detectors both in standard conditions and under in the influence of compression for GAN generated fake images. Jin et al. [20] designed a novel method based on visual and statistical image features implemented on real-world Sina Weibo dataset for microblog news verification.

**2.3 Multi-modal Approaches**

The modalities of text and image, when used in conjunction, mislead people by spreading fake news enormously. Although there are not many standard open-source datasets available for multi-modal fake news classification, having text and images, but few researchers have utilized the rarely available datasets for the task. Yang et al. [21] used explicit and implicit features from text and images using Convolutional Neural Networks (CNNs) for counterfeit news detection. Wang et al. [22] have created an end-to-end model that uses an event discriminator to classify fake news exploiting modalities of text as well as image. Their model uses CNN for text and VGG-19 to elicit photographic features and concatenating them to discriminate events and classify false news. Inspired by EANN, Khattar et al. [23] proposed a comparable framework using bi-LSTM instead of Text-CNN and formulated an architecture of auto-encoder and decoder. Same latent vectors were used for both encoder and decoder. Along with textual sentiment analysis and image segmentation process, Shah and Kobti [24] implemented cultural algorithm with situational and normative knowledge for multi-modal feature extraction. The final feature vector is passed through SVM classifier layer for veracity analysis experimented on Weibo and Twitter datasets. A neural network based multi-modal fake news detection system is proposed by Giachanou et al. [25] that combines textual, visual and sematic information being implemented on MediaEval , Politifact and GossipCop datasets. Vishwakarma et al. [26] processed multi-modal data in a different format by extracting the text written on images and then web scrapping this text claim in order to a reality parameter based on top 15 google search results. The value of the calculated reality parameter is compared against a threshold, which if exceeds a set point, the information is categorized as real otherwise fake. Meel and Vishwakarma [27] researched a framework for multi-modal fake news detection based on ensemble of Hierarchical Attention Network, Image Captioning and Forensics.

**2.4 Propagation Pattern Analysis**

The way rumour and fake news diffuse in a network is also a prominent feature of analysis and mitigation of misinformation. Non-sequential propagation structure of microblog posts is being identified by Ma et al. [28] to acquire discriminative attributes for generating powerful top-down and bottom-up representations using Recursive Neural Networks for rumour identification. They also proposed a kernel-based method of rumour detection termed as propagation tree kernel [29] to capture higher-order features for segregating various types of rumours based on the structure of their dissemination trees. Vu and Jung [30] proposed a propagation graph embedding method for rumour detection based on Graph Convolutional Neural Network. Experimental results illustrate the efficacy of their suggested method by reducing the detection error up to 10% as equated with state-of-the-arts. Other studies of combining attention mechanism with propagation graph structure [31] , an ensemble of user representation learning with news propagation dynamics [32] , implications of searchability on rumour self-correction [33] are also worth considering.

**2.5 Source Identification**

Propagation of fake news can be stopped up to an extent by identifying the source. Researchers have formulated various approaches to identify fake accounts on online platforms. In [34], authors have suggested a solution to discriminate fake Twitter accounts from genuine ones using dimensionality reduction and feature selection techniques. The authors in [35] use basic machine learning algorithms to classify fake accounts on the social media platform Instagram. It is not always viable to stop the propagation of fake news by recognizing and blocking the fake accounts because a user can create multiple accounts on an online platform or sometimes real accounts could also be a source of fake news. Louni and Subbalakshmi [36] addressed the problem of finding the origin of rumor in large scale social network. The salient feature of their proposed algorithm is probabilistically varying internode relationship strengths achieved by assigning random non-homogeneous edge weights to the original network graph. Query response method [37] using simple batch querying and interactive querying with directions is also out of the box analytical approach for rumor source detection.

## 2.6 Other Approaches

Authors in [38] propose a browser extension "BRENDA" implemented with Google Chrome browser for automated credibility assessment of claims using deep neural network. This automatic process is fast, real time, produces the evidence along with the classification and does not need to leave the web page. A Clickbait Video Detector (CVD) is designed by Varshney and Vishwakarma [39] to detect the clickbait videos circulating on web based on cognitive evidences extracted out of user profile, video content and human consensus. To extract the hidden patterns from unlabeled data a semi-supervised approach based on the temporal ensembling method is proposed by Meel and Vishwakarma [40] . Analysis of misinformation detection under varying time constraints under the consideration of three of different types of attacks: Evasion attacks, Poison attacks, Blocking attacks is thoroughly studied by Horne et al. [41].

## 2.7 Motivation and Research Gap

The detailed literature survey done in previous section and shortcomings of the existing systems encourages us to fuse transformer-based self-attention and fine-tuned transfer learning into a combined framework for veracity analysis of multi-modal online information. The key highlights of motivation and research gap are as follows:

- The main advantage of transformer models is that they are non-sequential, meaning that they don't require that the input sequence be processed in the order. This innovation allows transformers to be parallelized and scaled much more than previous NLP models.

- The Transformer supports multiple folds of parallelization and can achieve a new benchmark in terms of quality and performance.

- Transformer-based approaches BERT & ALBERT architecture are solely based on self-attention also called intra attention mechanism dispensing with recurrence and convolutions completely.

- Attention mechanism without recurrence and convolutions allows to draw global dependencies between input and output which removes all restrictions of sequential processing and motivates more of exploring the benefits of parallelization.

- BERT & ALBERT are bidirectionally trained language model supports a higher level of understanding about language structure and flow than single directional language model.

- Inception-ResNet-v2 pre-trained on ImageNet is being further fine-tuned for our veracity analysis task to harness the advantages of transfer learning which is more suitable for smaller datasets also and saves a lot of time of training the model from scratch.

- Motivation of using transfer learning is that the pre-trained models can be scaled for a variety of application specific tasks just by adding few final layers and fine tuning the weights as well as adjusting the hyperparameters on a comparatively smaller labelled dataset.

- Less work is reported in literature to address multi-modal information authenticity issue that will take the benefits of latest language modelling and transfer learning technologies.

- This work is expected to prove a landmark within the province of veracity analysis and encourage other researchers to propose further advancements to counter multiple visual and textual forgery formats that would take the scope of the solution to new heights.

**2.8 Problem Statement**

To better explain our work, we can mathematically formulate the problem statement as follows:

Let D is a set of training instances and n is the size of training dataset of news items.

$$D: \{x_1, x_2, \ldots \ldots, x_n\}$$

Each item, $x_i$ in set D contains the information $x_i: \{t_i, b_i, im_i, l_i\}$. Where,

$t_i \equiv title\ of\ the\ news\ item, x_i$

$b_i \equiv body\ of\ the\ news\ item, x_i$

$im_i \equiv image\ of\ the\ news\ item, x_i$

$l_i \equiv label\ of\ the\ news\ item, x_i$

$and, l_i \epsilon\ \{fake, real\}$

The prime aim is to predict the label $l_j$ of a news item $x_j: \{t_j, b_j, im_j\}$.

## 3. Proposed Multi-modal Architecture

It has become significant to incorporate images, as nearly all news, fake or real is accompanied by some visual data. A combination of Deep Learning Neural Network Models is used for veracity analysis of multi-modal data. We have used two modalities of textual and visual features in conjunction to classify a piece of news as fake or real. We use implicit features obtained from pre-trained models to train our custom neural network to obtain the final predictions using two independent architectures of Late Fusion and Early Fusion. We have used Inception-ResNet-v2 to extract visual features. The models BERT and ALBERT have been used to elicit textual attributes. Diverse forms of text input, like English articles, Chinese articles and Tweets have been used to make our model robust and usable across multiple platforms. The architecture of Multimodal Early Fusion and Late Fusion has been detailed in Figures 3 and 4 as well as the working steps are described in Algorithm 1 and 2 respectively. The frameworks and technical specifications used are defined in detail in the sub-sections below.

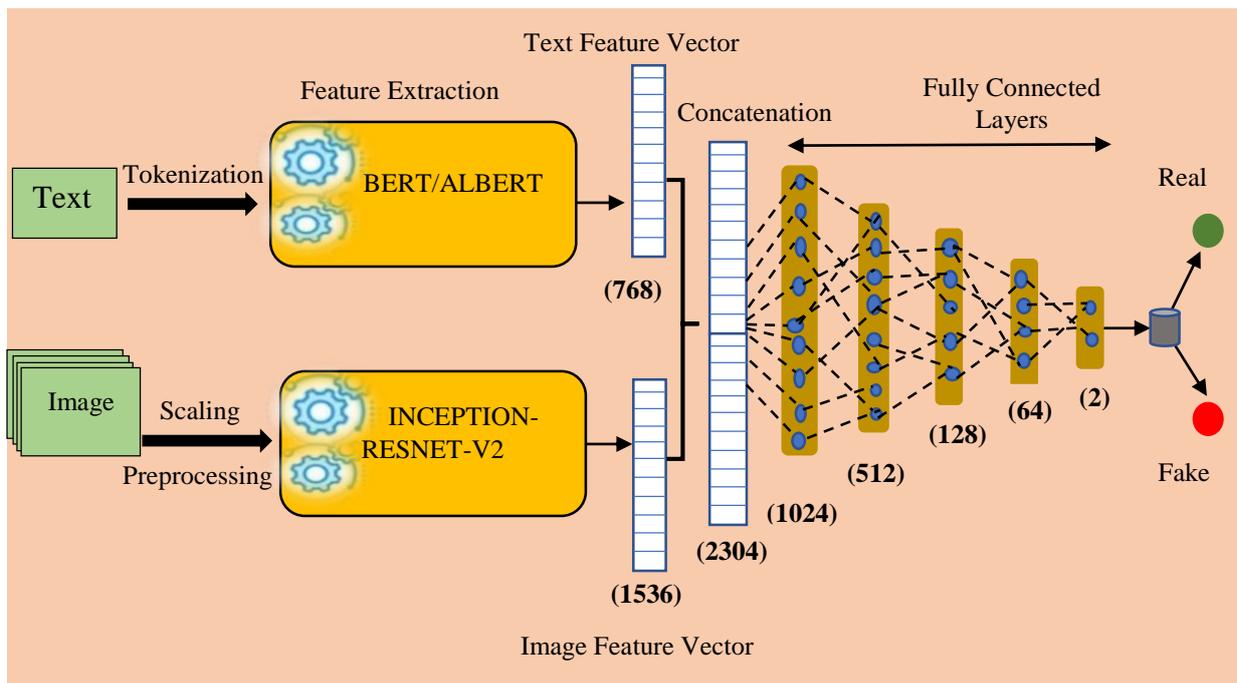

Figure 3: Architecture of Multimodal Early fusion

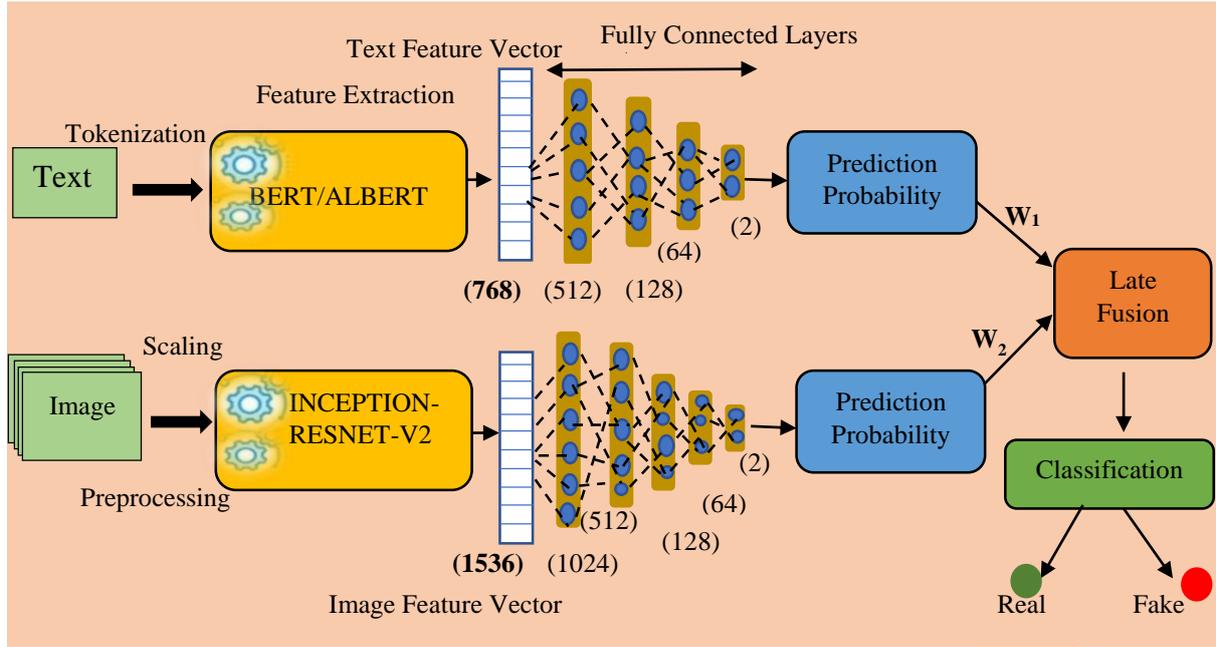

Figure 4: Architecture of Multimodal Late fusion

---

**Algorithm 1: Multimodal Early Fusion**

**Parameter Initialization**
- Input  $A = \{a_1, a_2, \ldots a_n\}$ is set of text vectors, $B = \{b_1, b_2, \ldots b_n\}$ is set of corresponding images, $C = \{c_1, c_2, \ldots c_n\}$ is set of labels for A and B.
- n is the size of training set
- Split A, B and C into three subsets for 70% training, 10% validation and 20% testing.

1: **For 1 to 30 epochs do**
2:     Extract feature vector using BERT/ALBERT model $M_1$ for text input A.
3:     Extract feature vector using Inception-ResNet-v2 model $M_2$ for image input B.
4:     Convert feature vectors obtained from $M_1$ and $M_2$ into unidimensional.
5:     Perform concatenation of feature vectors.
6:     Add a series of fully connected layers (dense, batch normalization, relu, dropout=0.4)
7:     Apply binary sigmoid classifier and calculate prediction probabilities
8:     Apply binary cross-entropy loss and Adam optimizer
9: **end for**
10: Evaluate the performance on test set

---

**Algorithm 2: Multimodal Late Fusion**

**Parameter Initialization**
- Input  $A = \{a_1, a_2, \ldots a_n\}$ is set of text vectors, $B = \{b_1, b_2, \ldots b_n\}$ is set of corresponding images, $C = \{c_1, c_2, \ldots c_n\}$ is set of labels for A and B.
- n is the size of training set
- Split A, B and C into three subsets for 70% training, 10% validation and 20% testing.

1: **For 1 to 30 epochs do**
2:     Extract feature vector using BERT/ALBERT model $M_1$ for text input A
3:     Add fully connected layers (dense, batch normalization, relu, dropout=0.4) to text feature vector
4:     Apply binary sigmoid classifier and calculate prediction probability $p_1$
3:     Extract feature vector using Inception-ResNet-v2 model $M_2$ for image input B
4:     Add fully connected layers (dense, batch normalization, relu, dropout=0.4) to image feature vector
5:     Apply binary sigmoid classifier and calculate prediction probability $p_2$
6:     Perform final prediction using weighted averaging late fusion ($\sum_{i=1}^{2} w_i p_i / \sum_{i=1}^{n} w_i$)
7:     Apply binary cross-entropy loss and Adam optimizer

| |
|---|
| 8: **end for** |
| 9:   Evaluate the performance on test set |

### 3.1 Pre-processing

The multi-modal veracity analysis framework is implemented on three datasets with textual data and image URLs to access the associated images. URLs were loaded using the Beautiful Soup library to access the images. Some of the images were inaccessible, which subsequently reduced the dataset as the corresponding rows were removed. Any attribute having null was handled by removing the corresponding rows. The dataset after all the mentioned processing was saved in the local drives for later access.

As far as preprocessing is concerned, textual data was not preprocessed as the model used was BERT and ALBERT which handles all types of words by its huge vocabulary size. The maximum sequence length for the BERT model for All data [42], Weibo [43] and MediaEval 2016 [44] datasets was fixed at 512, 400 and 20 respectively. The maximum sequence length was decided by taking the limit such that 95-98% of the word length comes under its range. For the dataset having tweets, preprocessing was performed by removing emojis, hyperlinks and tweet mentions from the textual part.

Images for every dataset were resized to (299,299) as this is the default size required by the Inception-ResNet-v2 model for feature extraction. All the images for every dataset were first scaled and converted to NumPy arrays. Then, minor preprocessing was done by using the Keras predefined function to obtain final images for input to our model. Features from textual data are extracted by using BERT and ALBERT language models based on the bidirectional encoding property of transformers and visual features are extracted by using the Inception-ResNet-v2 deep neural pre-trained model. After the feature vectors are obtained from both the modalities, separate standard scaling of both is done by using Sci-Kit Learns module "Standard Scaler".

### 3.2 BERT

BERT stands for Bidirectional Encoder Representations from Transformers. It was developed by research scientists working at Google AI Language [9]. It can perform various tasks including NLP, Next Sentence Prediction, Question-Answering, Classification, etc. What sets BERT apart from the other similar models is its ability to apply the bidirectional training property of a Transformer, which is a famous attention model, efficiently to language modeling framework. The Transformer is an attention mechanism that is used to learn the relationship between contexts of different words in text data. Transformers generally comprise two distinct mechanisms —one is an encoder that can read the input text, and another is a decoder that can produce the task prediction. But since BERT is only concerned with generating a language model, the decoder mechanism is not required. BERT is employed for a broad range of language-related tasks. All of them can be successfully performed only by augmenting a small-layer to the standard model. Classification-related tasks, such as sentiment analysis are quite similar in operation to the Next Sentence classification. We only need to add a classification layer on top of the output of the Transformer for the [CLS] token. The input representation for a token in BERT model is created by summating the equivalent token, segment and position embeddings. An apprehension of this structure is highlighted in figure 5.

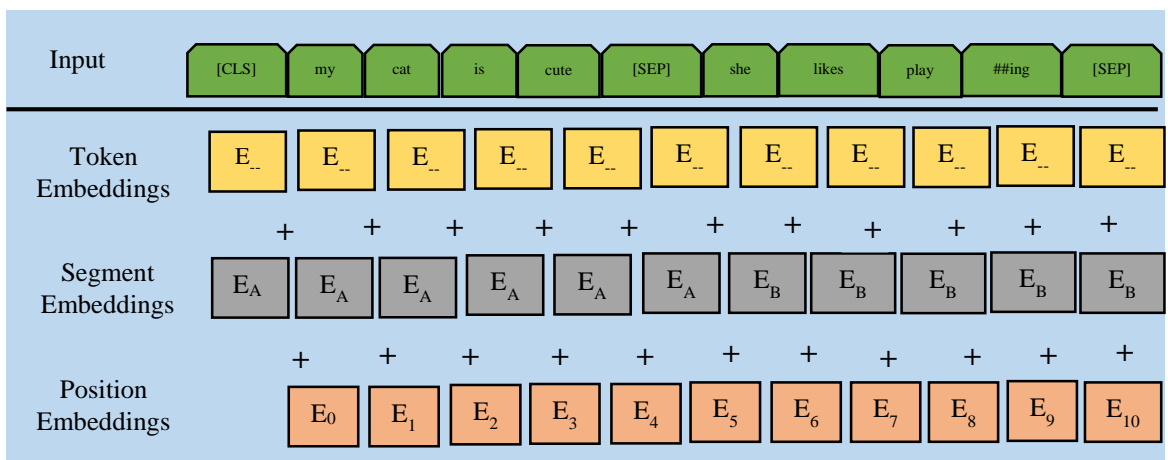

Figure 5: BERT Input Representation

Textual features associated with our dataset were obtained using BERT pre-trained model which was accessed from TensorFlow-Hub. BERT-base version was used for feature extraction as it was found to be suitable for our dataset. BERT-base extracts feature using attention mechanism by several encoding layers. Scaled Dot product attention and Multi Head attention detailed in figure 6 (a) and (b) are the core functions of BERT encoder. An attention function can be defined as mapping a query and a set of key-value pairs to an output, where the query, keys, values and output are all vectors. In scaled dot product attention, the input comprises of keys and queries of dimension $d_k$ and values of dimension $d_v$. To obtain the weights on the values the dot product of the query with all keys is divided by $\sqrt{d_k}$ and passed on to a SoftMax function as in equation (1). The set of queries, keys and values are arranged together into matrices Q, K and V. Multi-head attention can be described mathematically as in equation (2) and (3) .

$$Attention\ (Q, K, V) = softmax\left(\frac{QK^T}{\sqrt{d_k}}\right)V \qquad (1)$$

$$MultiHead(Q, K, V) = Concat(head_1, \ldots\ldots, head_h)W^O \qquad (2)$$

$$head_i = Attention(QW_i^Q, KW_i^K, VW_i^V) \qquad (3)$$

The projections are parameter metrices $W_i^Q \in \mathbb{R}^{d_{model} \times d_k}$, $W_i^K \in \mathbb{R}^{d_{model} \times d_k}$,

$W_i^V \in \mathbb{R}^{d_{model} \times d_v}$ and $W^O \in \mathbb{R}^{hd_v \times d_{model}}$ where h is the number of parallel attention layers or heads and $d_{model}$ is the dimension of each head. Twelve encoding layers extract implicit features from the text by taking input as input_ids, input_masks, segment_ids. BERT pre-trained model has an associated tokenizer which is built on large-sized vocabularies. In the Weibo dataset, character level tokenization is required as it is a Chinese language dataset where each character represents a word. For every sentence [CLS] and [SEP] tokens are appended, where CLS signifies Classification and SEP signifies special separating token.

- Input_word_ids: Encoded tokens using BERT-tokenizer
- Mask_ids: Separates useful and padded tokens (0 or 1)
- Segment_ids: Useful for pairwise training of sentences

These are tokenization inputs to the feed-forward BERT model which has a transformer architecture with 12 attention layers where each layer extracts features by attention mechanism to give output. The model is kept non-trainable to extract the bottleneck features which are obtained by freezing the weights of the entire model. BERT-base gives pooled output and sequence-output and pooled output is used for the further classification task. The vector length associated with everyrow is 768 sized one-dimension vector.

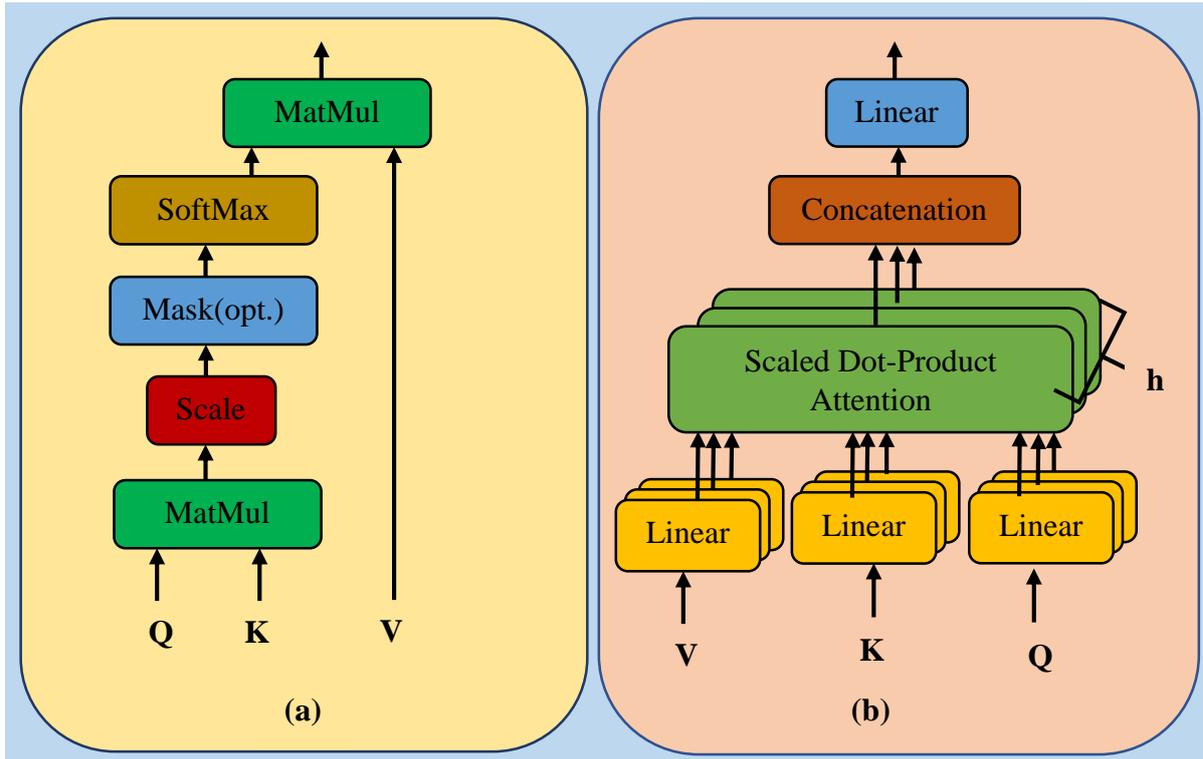

Figure 6: (a) Scaled Dot-Product Attention (b) Multi-Head Attention

In our proposed frameworks we harness the capabilities of BERT's multi-layer bidirectional Transformer encoder-based architecture grounded on the creative implementation of Transformer detailed in [45] by Vaswani et. al. The model is first initialized with the pre-trained parameters and then the last layer is replaced with a customized network to fine-tune the parameters with our labelled training datasets for the downstream task of binary fake news classification.

### 3.3. ALBERT

ALBERT, which stands for A Lite BERT [10], is a self-supervised learning model that was presented by Google researchers along with some upgrades to the BERT model. They came up with three major innovations that made it an even more refined model than BERT.

The foremost of them is **factorized embedding parameterization**. This played a major role for efficient allocation of the model's capacity. The size of the hidden layers was isolated from the vocab-embeddings size by using one-hot vectorization into first, an embedding space with lower dimensionality, and then to a hidden space altogether. Because of this, it was possible to increase the size of the hidden layer without much tuning of vocab-embeddings' parameter size. Another key feature was the **cross-parameter sharing**. This facilitated the sharing of parameters across all layers, keeping the depth of the model in control. Hence,

ALBERT has 18 times lesser parameters than BERT. Lastly, they introduced the concept of **inter-sentence coherence loss**. The original BERT model was not very reliable when it came to the next sentence prediction tasks. The introduction of SOP loss to model this inter-sentence coherence in ALBERT made it possible to increase the performance in such tasks.

### 3.4. Inception-ResNet-v2

Combining the best features of both the Inception and ResNet models, Google had proposed the Inception-ResNet-v2 [11].Inception-ResNet-v2 is a type of CNN (convolutional neural network) that is trained on the ImageNet dataset that has more than a million images. The network has a depth of 164 layers and can correctly classify images into about 1000 distinct groups such as pen, tree, box and many fruits. Henceforth, the network has learned sufficient feature representation techniques for a large set of images. The network has a standard size for input images, that is, 299x299.

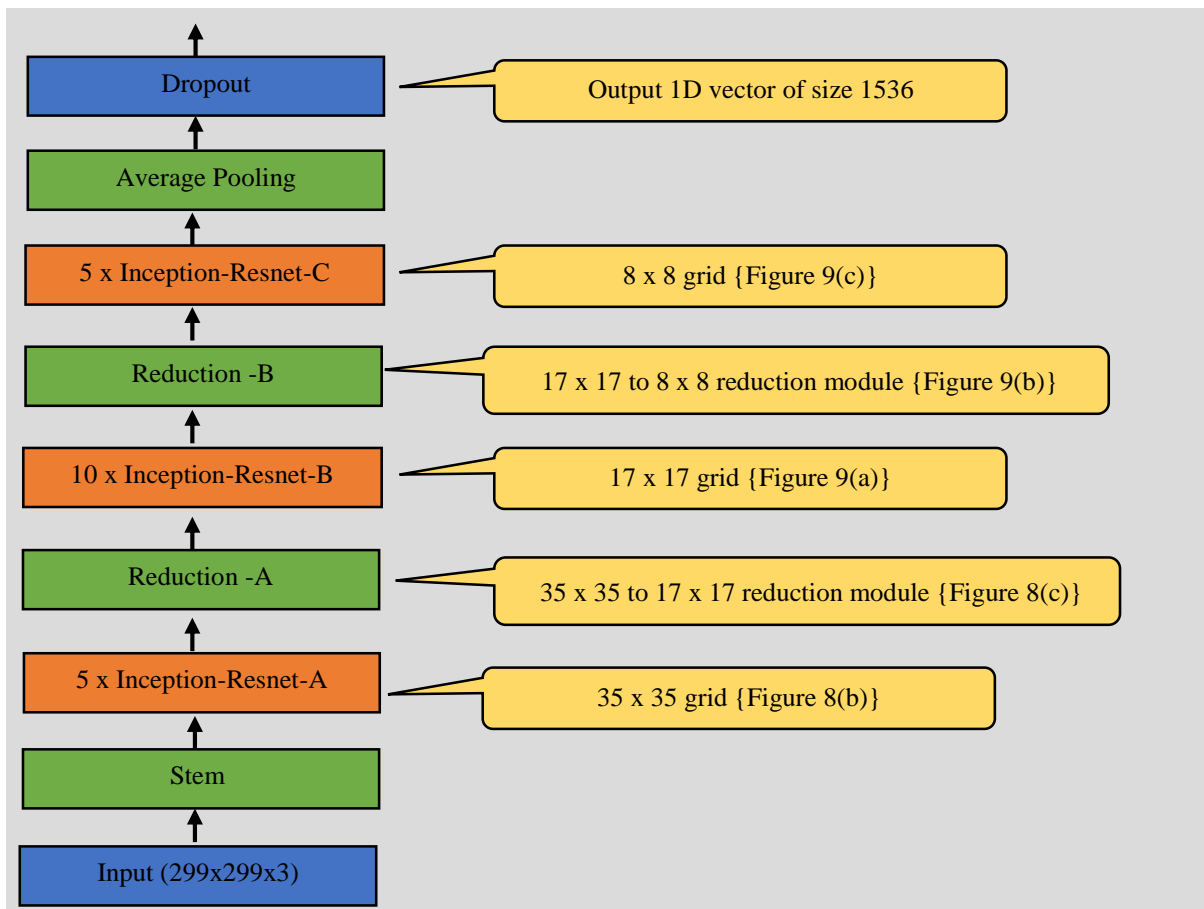

Figure 7: Schema for Inception-ResNet-v2 Network

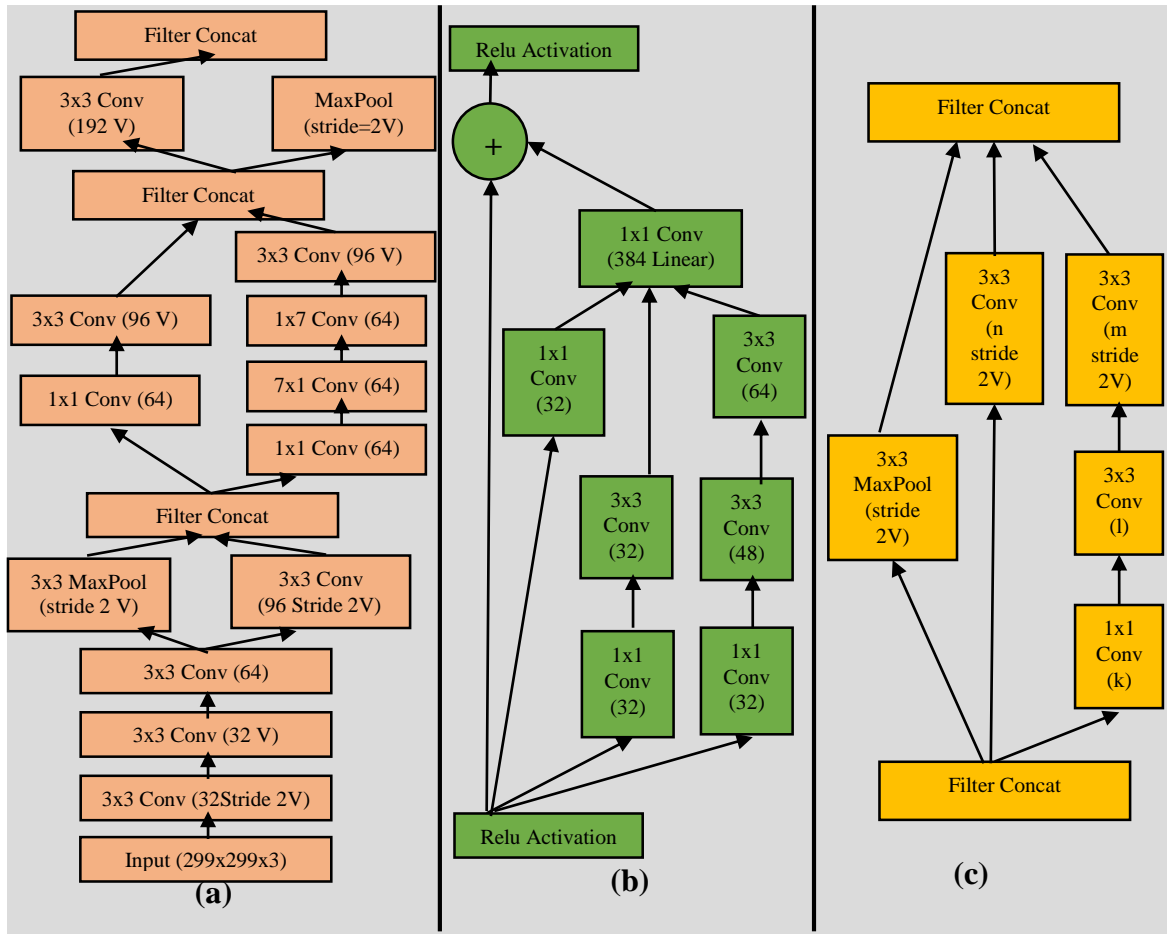

Figure 8: Schema for (a) Stem (b) Inception-Resnet-A (c) Reduction-A blocks of Inception-ResNet-v2 Network

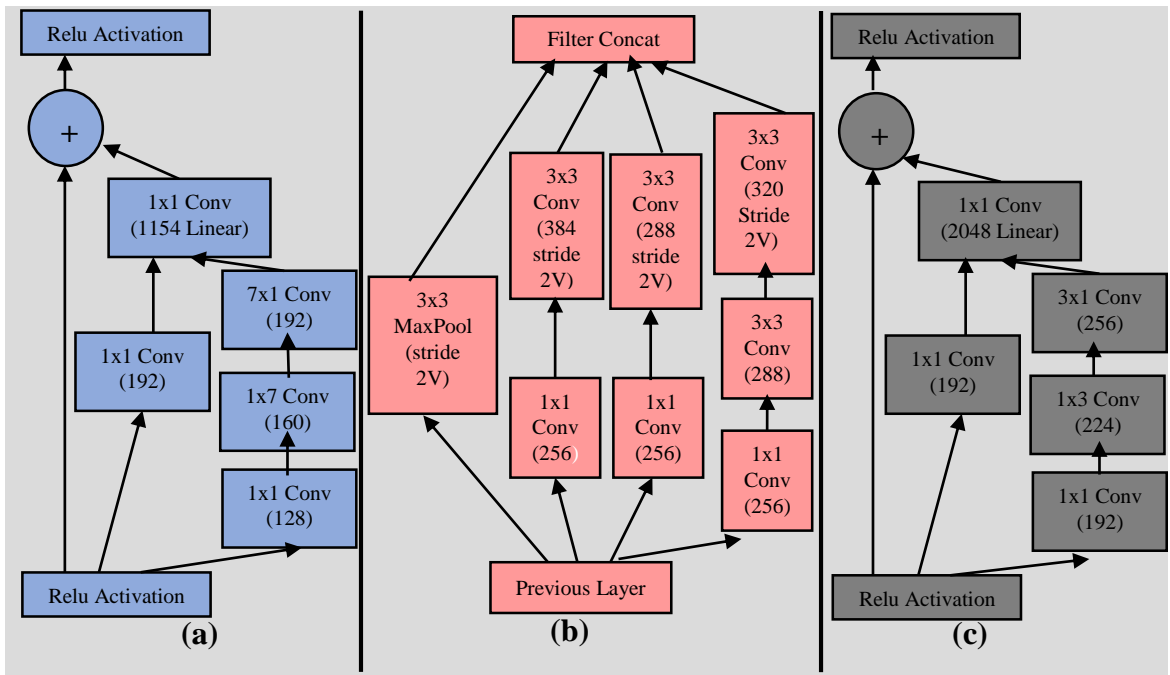

Figure 9: Schema for (a) Inception-Resnet-B (b) Reduction-B (c)Inception-Resnet-C blocks of Inception-ResNet-v2 Network

Figure 7 represents the large-scale schema of Inception-ResNet-v2 network. Figure 8(a), (b), (c) and Figure 9 (a), (b), (c) characterizes the comprehensive structure of its components. All the convolutions not marked with "V" in the figures are same-padded and convolutions marked with "V" are valid padded. The k, l, m and n represent the number of filters of the Reduction-A module for Inception-ResNet-v2 network and their values are 256, 256,384 and 384 respectively.After preprocessing the images, the InceptionResnetv2 model is loaded from the Keras Applications Library. After the model is loaded the final prediction layer is removed which classifies the image to 1000 available classes. The output of the penultimate layer is taken as our feature vector. The vector obtained from the InceptionResNetv2 model is a one-dimensional vector of size 1536.

Inception-Resnet model introduced the concept of residual connections that augment the output produced by the convolution operation of the basic inception model, to the input image. A necessary condition for this residual connection to work, the input image representation and the output produced after the convolution operation should have the same dimensions. Therefore, 1x1 convolutions are used after the original convolutions are performed, to keep the same depth. It is to be noted that after the convolution operation, the depth gets increased. InceptionResNet-v2 model is famous for attaining higher accuracy even at lower epoch cycles.

### 3.5 Multi-modal Fusion

Multi-modal fusion is the mechanism of combining information from diverse data channels into a single unit. According to the basic architectural levels, there are three basic approaches of fusion: recognition-based, decision based and hybrid multi-level fusion. The recognition-based fusion also popularly known as early fusion merge the feature vectors of each stream by means of integration (concatenation) mechanism. The main requirement of integrating feature vectors from multiple varied modalities is that their dimensions must be identical and this can be achieved by reshape operator. The decision-based fusion, popularly known as late fusion is the process of merging the final probability decisions coming from each branch. Hybrid multi-level fusion comprises of integrating the input modalities by distributing among the decision and recognition levels.

To perform a multi-modal fusion of independent text and image branches, first each of the modality is separately used to train two different models:

- A BERT/ALBERT model followed by custom layers is used to train the text model.

- A pre-trained Inception-ResNet-v2 followed by custom layers is used to train the image model.

Early fusion or recognition-based fusion is achieved by concatenating the feature vectors after converting them into identical dimensions. Late fusion or decision-based fusion is achieved by weighted averaging of the probabilities obtained by the independent predicted arrays from each parallel branch. The overall prediction is performed according to equation (4) by using two normalized weights ($w_1$, $w_2$); where $p_i$ is the probability of $i^{th}$ feature and $w_i$ is the corresponding assigned weight. The weights assigned for text classifier is $w_1$ and for image classifier is $w_2$.

$$Final\ prediction = \frac{\sum_{i=1}^{2} w_i p_i}{\sum_{i=1}^{2} w_i} \qquad (4)$$

## 4. Implementation Details

In this sub-section, we extensively detail the experimentation settings, datasets, parameter selection and insights gained from the wide range of experiments performed. The implementation is done on Google Colab which offers up to 13.53 free RAM and 12 GB NVIDIA Tesla K80 GPU. The proposed framework is built and implemented in Python 3 on top of the Keras deep learning framework. All the three datasets are split in the ratio of 7:1:2 for the train, validate and test clusters respectively to make our experimentation better and effective. The performance scores and comparison of the proposed framework are listed in terms of F1-measure, accuracy, recall, precision evaluation metrics. Analysis of the results is being done in numerical as well as in graphical representation using the shapes of accuracy-loss trends with epochs and area under the curve plots individually for each dataset.

### 4.1 Datasets

To substantiate the efficacy of our proposed framework we experimented with three different datasets comprising unique properties of language and writing style. The first dataset "All Data" contains English news articles, the second one "Weibo" contains Chinese news articles and the third one twitter dataset "MediaEval 2016" comprising of tweets on various topics. All three datasets have text as well as images. The following Table1 highlights the details of the datasets used.

Table1: Dataset Details

| Dataset Name | Details | Attributes Used | Total entries | Fake news count | Real news count |
|---|---|---|---|---|---|
| All Data [42] | Dataset contains English News Articles with 54 attributes | Title, Text, Image, Label | 20015 | 11941 | 8074 |

| | including site URL, Image URL, Body, Headline, Label | | | | |
|---|---|---|---|---|---|
| Weibo [43] | Dataset contains Chinese News articles with four columns title, text, image URL and label | Title, Text, Image, Label | 5250 | 2767 | 2438 |
| MediaEval 2016 [44] | Dataset contains tweets related to multiple events with post_id, post_text, user_id, image_ids, username, timestamp and label | Tweet text, Image, Label | 15519 | 6875 | 8644 |

### 4.1.1 All Data

All Data dataset has news articles in English language. It contains title, title-length, body/text, text_length, site_url, sentiment words count, title-length, image_url, and other metadata of a total of 54 attributes for almost 25000 news articles with binary labels of real and fake. The attributes which were used are the title of the news post, the body of the post, associated image and label. Cleaning and preprocessing were performed on the dataset before use. Entire rows that had null values for any of these four attributes were removed. Also, entire rows were removed if the image from image_url could not be fetched from the given source. The Title and Text attributes were concatenated into a single attribute for each row. After cleaning, the dataset had 20,015 rows of data with 11 914 fake news and 8 047 real news count.

### 4.1.2 Weibo Dataset

Weibo dataset is the standard Chinese dataset having microblogs from the Chinese microblogging site Sina Weibo. It is one of the most popularly used dataset and has news articles in the Chinese Language (Mandarin). News articles are accompanied by an image URL, from which images were fetched and stored before use. The dataset contains title, text, image and binary label (Real/Fake) associated with each news article. Entire rows that had null values for any of these four attributes were removed. Also, entire rows were removed if the image from image_url could not be fetched from the given source. After initial preprocessing, the dataset contains 5250 total news articles out of which 2767 are fake and 2438 are real news articles. The Title and Text attributes were concatenated into a single attribute for each row. For preprocessing, tokenization was performed using a Chinese character level tokenizer. BERT model has a tokenizer which is the most suitable for this task. Tokenization was done on a character level as the Chinese language is to character in the same way as the English language is to words. After this, the dataset was ready to be fed as input to our model.

### 4.1.3 MediaEval 2016

This dataset is part of the MediaEval task of 2016 and is available online at the Image Verification corpus [44]. The dataset contains tweets related to multiple events. It has been extensively used since Twitter is amongst the most commonly used social media platforms and is highly susceptible to the creation as well as propagation of fake news. It has several attributes: post_id, post_text, user_id, image_ids, username, timestamp and label. The attributes which were used are Post_text - The body of the post, Image_ids – link to an image related to the post and label - Classified as real or fake. Cleaning and preprocessing were performed on the dataset before use. Entire rows that had null values for any of these three attributes were removed. Also, entire rows were removed if the image from image_ids could not be fetched from the given source. After cleaning the dataset contains a total of 15 519 tweets with images and binary labels out of which 6875 tweets are fake and 8644 tweets are real.

### 4.2 Model Parameter Description

The network is being trained and tested for three datasets. BERT-base and ALBERT-base pretrained networks are used for text feature extraction and Inception-ResNet-v2 pretrained model is used for image feature vector extraction. All the model parameters are kept identical for these three datasets except the maximum sequence length which is set to 512, 400 and 20 respectively for All Data, Weibo and MediaEval dataset. The network is trained with Adam optimizer having learning rate 0.0001, momentum parameter $β_1 = 0.9$ and $β_2=0.980$ up to 30 epochs with a batch size of 128. To train Early Fusion architecture text and image feature vectors of length 768 and 1536 are concatenated together followed by five dense layers for smoother dimensionality reduction. The neurons in dense layers are 1024, 512, 128, 64 and 2 correspondingly. Two neuron final dense layer with sigmoid activation supports binary classification. Throughout the custom layers dropout probability is 0.4 with batch normalization. The initial four dense layers are equipped with ReLU activation function and loss is calculated with binary cross entropy loss. The Late fusion multi-modal architecture calculates independent prediction probability $p_1$ for text and $p_2$ for image stream. These probabilities are fused with weighted averaging late fusion framework analysed extensively for four different weight combinations ($w_1$, $w_2$) = {(0.4,0.6), (0.5, 0.5), (0.6,0.4), (0.7, 0.3)}. Feature vector extracted from text is followed by four fully connected dense layers of 512, 128, 64 and 2 neurons. Feature vector extracted from image part is followed by five fully connected dense layers of 1024, 512, 128, 64 and 2 neurons. All other hyperparameters are same for both the architectures to have an efficient analysis and fair comparison of the prediction preciseness.

### 4.3 Result Analysis

We employed performance measures like Accuracy, Precision, Recall, F1Score,

Accuracy-epoch curve, loss-epoch curve and ROC Curve to analyse and assess the outcomes attained by performing the experimentations on three datasets. The results are presented in Table 2, Table 3 and Table 4 for each dataset sequentially. On each dataset, we applied multi-modal early fusion and late fusion with four different weight combinations for weighted averaging the results using both variants of the framework i.e. BERT and Inception-ResNet-v2 as well as ALBERT and Inception-ResNet-v2. For a clear understanding and insights gained through these results, pictorial representation in terms of graphs is also presented.

The highest accuracy achieved on All Data dataset is 97.19% with ALBERT and Inception-ResNet-v2 framework. Table 2 details in with the outcomes of applying early fusion and late fusion on the dataset. Figure 10 (a), (b) and (c) concentrates on the accuracy versus epoch curve, loss versus epoch curve and ROC curve.

Table 2: Result analysis on All Data Dataset

| Fusion | Weightage of text & image streams | Accuracy (%) | Precision (%) | Recall (%) | F1-score (%) |
|---|---|---|---|---|---|
| BERT+Inception Resnet-v2 | | | | | |
| Early Fusion | Feature vector concatenation | 94.10 | 95.02 | 94.88 | 94.95 |
| Late Fusion | (w1, w2) = (0.4,0.6) | 85.75 | 78.33 | 82.17 | 80.20 |
| | (w1, w2) = (0.5,0.5) | 90.33 | 88.06 | 92.64 | 90.29 |
| | (w1, w2) = (0.6,0.4) | 95.76 | 92.20 | 97.45 | 94.75 |
| | (w1, w2) = (0.7,0.3) | 91.33 | 91.20 | 89.74 | 90.46 |
| ALBERT+Inception Resnet-v2 | | | | | |
| Early Fusion | Feature vector concatenation | 96.41 | 95.48 | 94.38 | 94.93 |
| Late Fusion | (w1, w2) = (0.4,0.6) | 87.06 | 85.77 | 89.23 | 87.47 |
| | (w1, w2) = (0.5,0.5) | 92.70 | 90.26 | 92.11 | 91.17 |
| | **(w1, w2) = (0.6,0.4)** | **97.19** | **97.00** | **99.00** | **97.99** |
| | (w1, w2) = (0.7,0.3) | 91.29 | 92.66 | 88.24 | 90.39 |

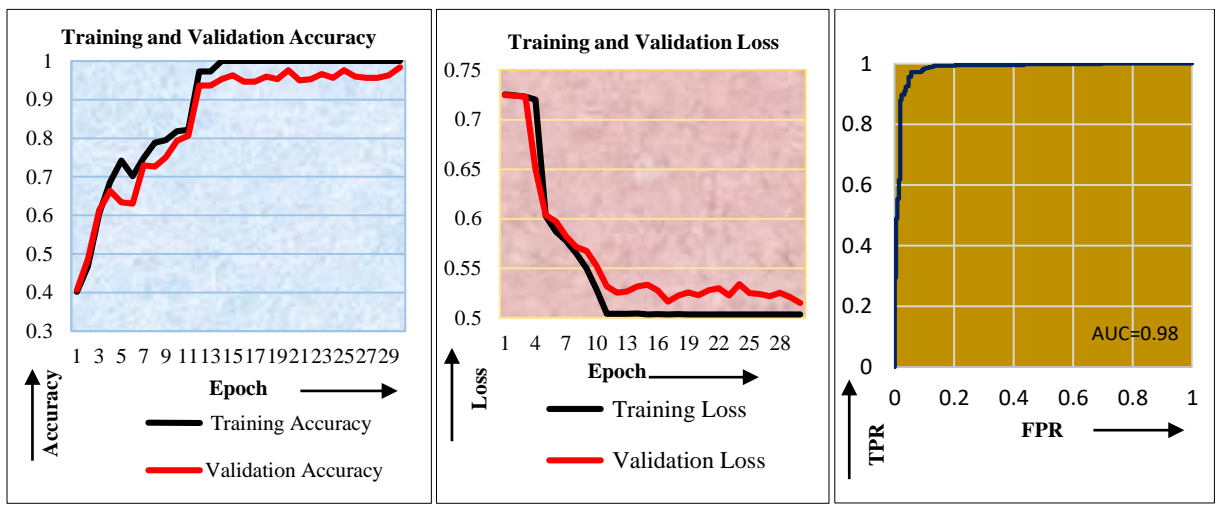

(a)          (b)          (c)

Figure 10: (a) Accuracy-Epoch curve (b) Loss-Epoch curve (c) ROC curve for All Data dataset

Table 3: Result analysis on Weibo Dataset

| Fusion | Weightage of text & image streams | Accuracy (%) | Precision (%) | Recall (%) | F1-score (%) |
|---|---|---|---|---|---|
| BERT+Inception Resnet-v2 | | | | | |
| Early Fusion | Feature vector concatenation | 89.61 | 78.43 | 76.34 | 77.38 |
| Late Fusion | (w1, w2) = (0.4,0.6) | 80.55 | 79.42 | 83.17 | 81.25 |
| | (w1, w2) = (0.5,0.5) | 83.40 | 86.27 | 78.90 | 82.42 |
| | (w1, w2) = (0.6,0.4) | 89.92 | 90.20 | 93.42 | 91.78 |
| | (w1, w2) = (0.7,0.3) | 82.45 | 87.32 | 74.65 | 80.49 |
| ALBERT+Inception Resnet-v2 | | | | | |
| Early Fusion | Feature vector concatenation | 91.00 | 91.27 | 89.79 | 90.52 |
| Late Fusion | (w1, w2) = (0.4,0.6) | 83.24 | 86.47 | 94.33 | 90.23 |
| | (w1, w2) = (0.5,0.5) | 83.67 | 87.24 | 92.22 | 89.66 |
| | **(w1, w2) = (0.6,0.4)** | **94.28** | **94.07** | **95.82** | **94.94** |
| | (w1, w2) = (0.7,0.3) | 88.04 | 87.00 | 90.22 | 88.58 |

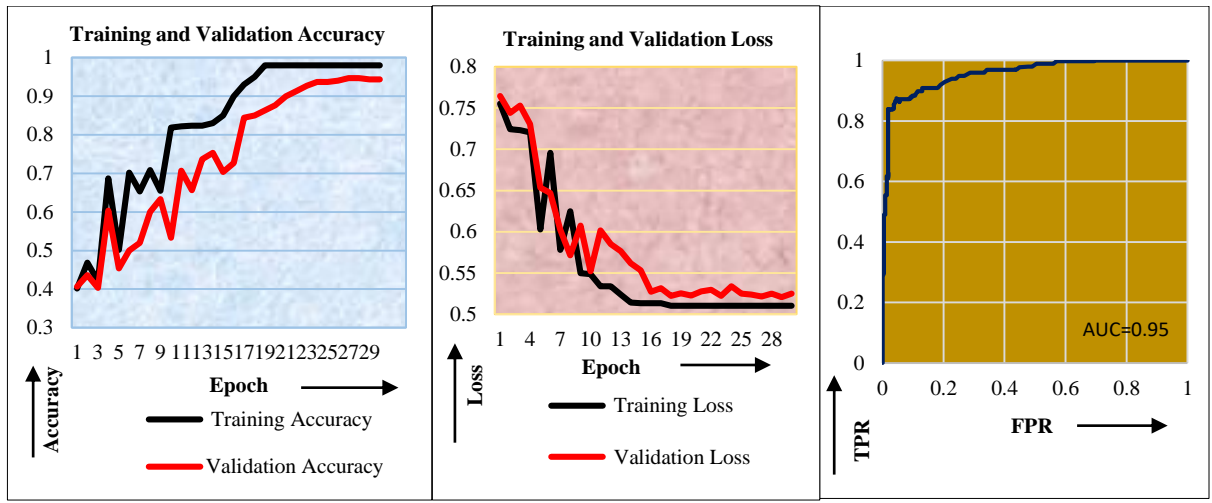

(a)          (b)          (c)

Figure 11: (a) Accuracy-Epoch curve (b) Loss-Epoch curve (c) ROC curve for Weibo dataset

Table 4: Result analysis on MediaEval Dataset

| Fusion | Weightage of text & image streams | Accuracy | Precision | Recall | F1-score |
|---|---|---|---|---|---|
| BERT+Inception Resnet-v2 | | | | | |
| Early Fusion | Feature vector concatenation | 66.69 | 59.52 | 40.86 | 48.46 |
| Late Fusion | (w1, w2) = (0.4,0.6) | 54.22 | 59.80 | 54.17 | 56.84 |
| | (w1, w2) = (0.5,0.5) | 59.67 | 68.52 | 52.86 | 59.68 |
| | (w1, w2) = (0.6,0.4) | 67.04 | 74.91 | 62.50 | 68.14 |
| | (w1, w2) = (0.7,0.3) | 61.51 | 61.27 | 65.77 | 63.44 |
| ALBERT+Inception Resnet-v2 | | | | | |
| Early Fusion | Feature vector concatenation | 67.79 | 60.30 | 76.00 | 67.24 |
| Late Fusion | (w1, w2) = (0.4,0.6) | 57.61 | 67.25 | 52.28 | 58.82 |
| | (w1, w2) = (0.5,0.5) | 69.07 | 70.86 | 52.47 | 60.29 |
| | **(w1, w2) = (0.6,0.4)** | **75.33** | **78.00** | **62.50** | **69.39** |
| | (w1, w2) = (0.7,0.3) | 63.01 | 74.52 | 52.09 | 61.32 |

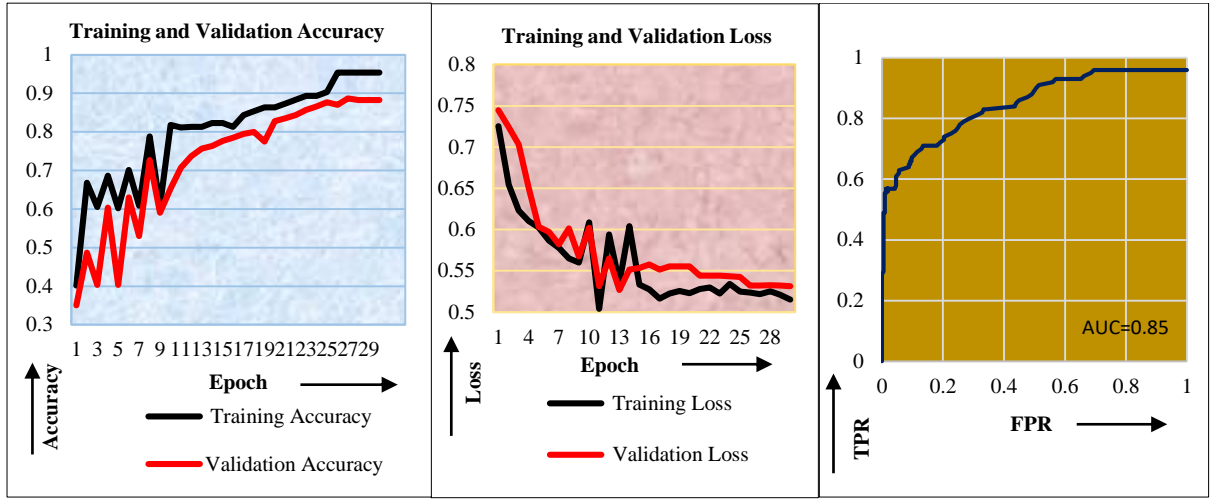

**(a)**            **(b)**            **(c)**

Figure 12: (a) Accuracy-Epoch curve (b) Loss-Epoch curve (c) ROC curve for MediaEval dataset

It is evident from Table 3 and Table 4 that highest accuracy achieved on Weibo and MediaEval dataset are 94.28% and 75.33% respectively. Figure 11 and Figure 12 elaborates the effectiveness of the training and validation process as well as the robustness of the framework in terms of ROC curve. The above discussion reinforced with Tables 2, 3, and 4 along with Fig. 10,11 and 12 verifies that our proposed multi-modal fusion architecture using fine-tuned self-attention and transfer learning accomplishes promising performance and outstanding precisions for veracity analysis of web information.

### 4.4 State-of-the-art Comparison

To compare the efficacy of our designed architecture with contemporary methods, we calculated the performance of five different baseline methods in terms of accuracy, precision, recall and F1 score with the dataset split as 7:1:2 for training, validation and testing. State-of-the-art juxtaposition on each of the All Data, Weibo and MediaEval datasets are outlined in Table 5, Table 6 and Table 7, correspondingly. The approaches used as baselines for state-of-the-art comparison are as follows:

- Vishwakarma et al. [26] implemented reverse search method of extracted text from images on internet to calculate a reality parameter by using top search results.
- Wang et al [22] devised multi-modal fake news detector by incorporating text CNN and VGG19 along with event discriminator module. To have a fair comparison, we removed the event discriminator module and compared on binary classifier with our datasets.

- Early fusion of textual features extracted by LSTM and visual features extracted by VGG19 is being proposed and implemented by Jin et al. [46].
- Lago et al. [47] proposed fake news detection multi-modal framework by a combination of classical and advanced techniques such as Mean-filter noise residue inconsistency, Splice buster, JPEG ghosts, TF-IDF, cosine similarity, Logistic Regression, CNN etc.
- Khattar et al. [23] designed variational autoencoder system with three core parts encoders, decoder and a binary fake news detection using Bi-LSTM and VGG19 methods for veracity analysis of online articles.

Table 5: State-of-the-art comparison on All Data Dataset

| Reference | Input Modality | Proposed Method /Classifier | Acc. (%) | P (%) | R (%) | F (%) |
|---|---|---|---|---|---|---|
| Vishwakarma et al. [26] | Text +Image | Rule based classifier, Reality Parameter | 88.00 | 87.90 | 88.80 | 88.34 |
| Wang et al. [22] | Text +Image | CNN + VGG19 | 84.27 | 86.72 | 83.28 | 84.96 |
| Jin et al. [46] | Text +Image | LSTM + VGG 19 | 79.66 | 87.19 | 79.61 | 83.22 |
| Lago et al. [47] | Text +Image | Classical and advanced image forensics, CNN, LR, RF, Jaccard's similarity, TF-IDF, cosine similarity | 92.34 | 93.06 | 94.06 | 93.56 |
| Khattar et al. [23] | Text +Image | BiLSTM + VGG19 | 85.70 | 87.81 | 83.03 | 85.34 |
| **Proposed Method** | **Text + Image** | **ALBERT + Inception-ResNet-v2** | **97.19** | **97.00** | **99.00** | **97.99** |

Table 6: State-of-the-art comparison on weibo Dataset

| Reference | Input Modality | Proposed Method /Classifier | Acc. (%) | P (%) | R (%) | F (%) |
|---|---|---|---|---|---|---|
| Vishwakarma et al. [26] | Text +Image | Rule based classifier, Reality Parameter | 90.88 | 91.24 | 89.16 | 90.19 |
| Wang et al. [22] | Text +Image | CNN + VGG19 | 82.73 | 84.70 | 81.22 | 82.92 |
| Jin et al. [46] | Text +Image | LSTM + VGG 19 | 78.80 | 86.20 | 68.60 | 76.40 |
| Lago et al. [47] | Text +Image | Classical and advanced image forensics, CNN, LR, RF, Jaccard's similarity, TF-IDF, cosine similarity | 80.02 | 79.66 | 83.24 | 81.41 |
| Khattar et al. [23] | Text +Image | BiLSTM + VGG19 | 82.41 | 85.42 | 76.93 | 80.95 |
| **Proposed Method** | **Text + Image** | **ALBERT + Inception-ResNet-v2** | **94.28** | **94.07** | **95.82** | **94.94** |

Table 7: State-of-the-art comparison on MediaEval Dataset

| Reference | Input Modality | Proposed Method /Classifier | Acc. (%) | P (%) | R (%) | F (%) |
|---|---|---|---|---|---|---|
| Vishwakarma et al. [26] | Text +Image | Rule based classifier, Reality Parameter | 73.45 | 75.20 | 71.00 | 73.04 |
| Wang et al. [22] | Text +Image | CNN + VGG19 | 71.50 | 82.20 | 63.80 | 71.84 |
| Jin et al. [46] | Text +Image | LSTM + VGG 19 | 68.20 | 78.00 | 61.50 | 68.77 |
| Lago et al. [47] | Text +Image | Classical and advanced image forensics, CNN, LR, RF, Jaccard's | 75.00 | 76.32 | 71.67 | 73.92 |

| | | similarity, TF-IDF, cosine similarity | | | | |
|---|---|---|---|---|---|---|
| Khattar et al. [23] | Text +Image | BiLSTM + VGG19 | 74.52 | 80.10 | 71.90 | 75.77 |
| **Proposed Method** | **Text + Image** | **ALBERT+ Inception-ResNet-v2** | **75.33** | **78.00** | **62.50** | **69.39** |

The above discussion concludes that our proposed model for veracity analysis of web information is reasonably promising. The accuracy over all the three datasets All Data, Weibo and MediaEval is a decent development over the parallel methods. Fine-tuned self-attention and transfer learning are the prominent technologies that have facilitated refining the preciseness of our results. Finally, merging multiple data streams with late fusion helps us achieve 97.19% highest fake news detection accuracy.

## 5. Conclusion and Future Works

With the exponential emergence of social media as a globally used platform for information circulation and social interactions in the recent decade, we have encountered massive escalation in fake news dispersion. Therefore, the need to establish baselines for information-assessment across all information flow channels has demanded the research community's attention. In this work, we have considered both the textual and visual characteristics of the input news instances. To use these features in veracity analysis, the latest pre-trained deep learning models BERT, ALBERT and Inception-ResNet-v2 have been used. These models are fine-tuned according to our application and dataset requirements to get precise results. Fine-tuning for leveraging existing knowledge in a deep learning model can be potentially beneficial for automatic detection and analysis of fake news. Extensive evaluation of the proposed Multimodal Early and Late fusion framework on three renowned datasets of English news articles, Chinese news articles, and Tweets demonstrated our proposed model's commendable performance in identifying the fake news articles. Its ability to work on large and diverse datasets makes the framework a strong contender for better fake news detection algorithms than the current ones in practice.

There is still a lot of scope for further innovation and future improvements in the proposed architecture. Deep Learning is an ever-evolving field with new and better algorithms being created now and then. Self-attention-based models leveraging the benefits of transformers, dispensing with convolutions and recurrence entirely, will be the future scope of a wide range of language modelling-based application areas. The Transformers can be progressed to figure out the solutions of input and output modalities other than text such as video, audio and images to investigate local, constrained attention mechanisms for handling massive inputs and outputs proficiently. Making processing less sequential could be another approaching research objective. We hope that this research would work as a baseline for future researchers who can further optimize this solution and make it more relevant with time. Eventually, the proposed work facilitates some reliable mechanism to help in the veracity analysis of web information and more importantly, apprehends the much-needed awareness that everything we read online is not always true.

There is no conflict of interest